\documentclass[%
 reprint,
 amsmath,amssymb,
 aps
]{revtex4-1}

\usepackage[english]{babel}

\makeatletter
\def\bbl@set@language#1{%
  \edef\languagename{%
    \ifnum\escapechar=\expandafter`\string#1\@empty
    \else\string#1\@empty\fi}%
  \@ifundefined{babel@language@alias@\languagename}{}{%
    \edef\languagename{\@nameuse{babel@language@alias@\languagename}}%
  }%
  \select@language{\languagename}%
  \expandafter\ifx\csname date\languagename\endcsname\relax\else
    \if@filesw
      \protected@write\@auxout{}{\string\select@language{\languagename}}%
      \bbl@for\bbl@tempa\BabelContentsFiles{%
        \addtocontents{\bbl@tempa}{\xstring\select@language{\languagename}}}%
      \bbl@usehooks{write}{}%
    \fi
  \fi}
\newcommand{\DeclareLanguageAlias}[2]{%
  \global\@namedef{babel@language@alias@#1}{#2}%
}
\makeatother

\DeclareLanguageAlias{en}{english}

\usepackage{graphicx}
\usepackage{amsmath}
\usepackage{array}
\usepackage{amssymb}
\usepackage{dcolumn}
\usepackage{bm}
\let\savecorresponds\corresponds
\let\corresponds\relax
\usepackage{mathabx}
\usepackage{enumitem} 
\usepackage{tabularx}
\let\corresponds\savecorresponds

\def\pd2v#1#2#3{\frac{\partial^2 #1}{\partial #2 \partial #3}}

\def \2x2mat#1#2#3#4{
\left( \begin{array}{cc}
#1 &  #2 \\  #3 &  #4
\end{array} \right)
}

\begin{document}

\preprint{APS/123-QED}

\title{Adaptive Quantum Optics with Spatially Entangled Photon Pairs}

\author{Hugo Defienne}
\email{defienne@princeton.edu}
\author{Matthew Reichert}%
\author{Jason W. Fleischer}%
\affiliation{%
Department of Electrical Engineering, Princeton University, Princeton, NJ 08544, USA
}%

\date{\today}

\begin{abstract}
Light shaping facilitates the preparation and detection of optical states and underlies many applications
in communications, computing, and imaging. In this Letter, we generalize light shaping to the quantum
domain. We show that patterns of phase modulation for classical laser light can also shape higher orders
of spatial coherence, allowing deterministic tailoring of high-dimensional quantum entanglement. By
modulating spatially entangled photon pairs, we create periodic, topological, and random patterns of
quantum illumination, without effect on intensity. We then structure the quantum illumination to
simultaneously compensate for entanglement that has been randomized by a scattering medium and to
characterize the medium’s properties via a quantum measurement of the optical memory effect. The results
demonstrate fundamental aspects of spatial coherence and open the field of adaptive quantum optics.
\end{abstract}

\maketitle

Light shaping is indispensable in many areas of optics. Examples range from the pioneering works of Gabor holography~\cite{gabor_new_1948} and Zernike phase masks~\cite{zernike_how_1955} to the recent breakthroughs enabled by spatial light modulators (SLMs)~\cite{yeh_optics_2009}. Static SLM patterns have led to advanced pulse shaping~\cite{weiner_femtosecond_2000}, super-resolution microscopy~\cite{klar_fluorescence_2000}, and 3D surface imaging, while dynamic patterns underlie video projection and adaptive optics. All of these methods can be used for, and be enhanced by, quantum illumination. Indeed, wavefront shaping has been used to manipulate orbital angular momentum modes (OAMs) of quantum light~\cite{leach_quantum_2010,fickler_quantum_2012} and to pre-compensate photon scattering in disordered media~\cite{defienne_two-photon_2016,wolterink_programmable_2016}, but to date structure has been imposed independently on each photon of an entangled pair. This restriction was due partly to the use of the other photon for heralding and partly due to the extraordinary difficulty of measuring higher-order spatial coherence. The manipulation has therefore been classical, as there is no substantial difference between shaping the wavefront of a single photon and that of coherent light; rather, true quantum control arises from shaping the correlations within the joint probability distribution. Here, we consider spatially entangled photon pairs and experimentally structure second-order spatial coherence across the entire biphoton distribution function. 

Intuition for quantum wavefront shaping follows from the generalized concept of optical coherence~\cite{glauber_quantum_1963}. First-order coherence of laser light allows intensity shaping by phase modulation of the angular spectrum. For second-order coherence, phase modulation acts on the two-photon wavefunction of spatially entangled photon pairs, which has repercussions on intensity correlations (i.e. coincidences) in the reciprocal space. This means that a given pattern on an SLM can be used to shape both classical and quantum light, as long as measurements are performed in their respective region of coherence. This correspondence is remarkable, as classical methods are exponentially easier to perform: their signal is higher and their measurements simpler. Classical control and feedback are thus exponentially quicker. In practice, optimization in the quantum domain can by bypassed, as quantum signals can piggyback on the classical parameters. The result is classical design for quantum resources, enabling the highest orders of performance with the lowest order of wavefront manipulation.

\begin{figure}
\centering
\includegraphics[width=1 \columnwidth]{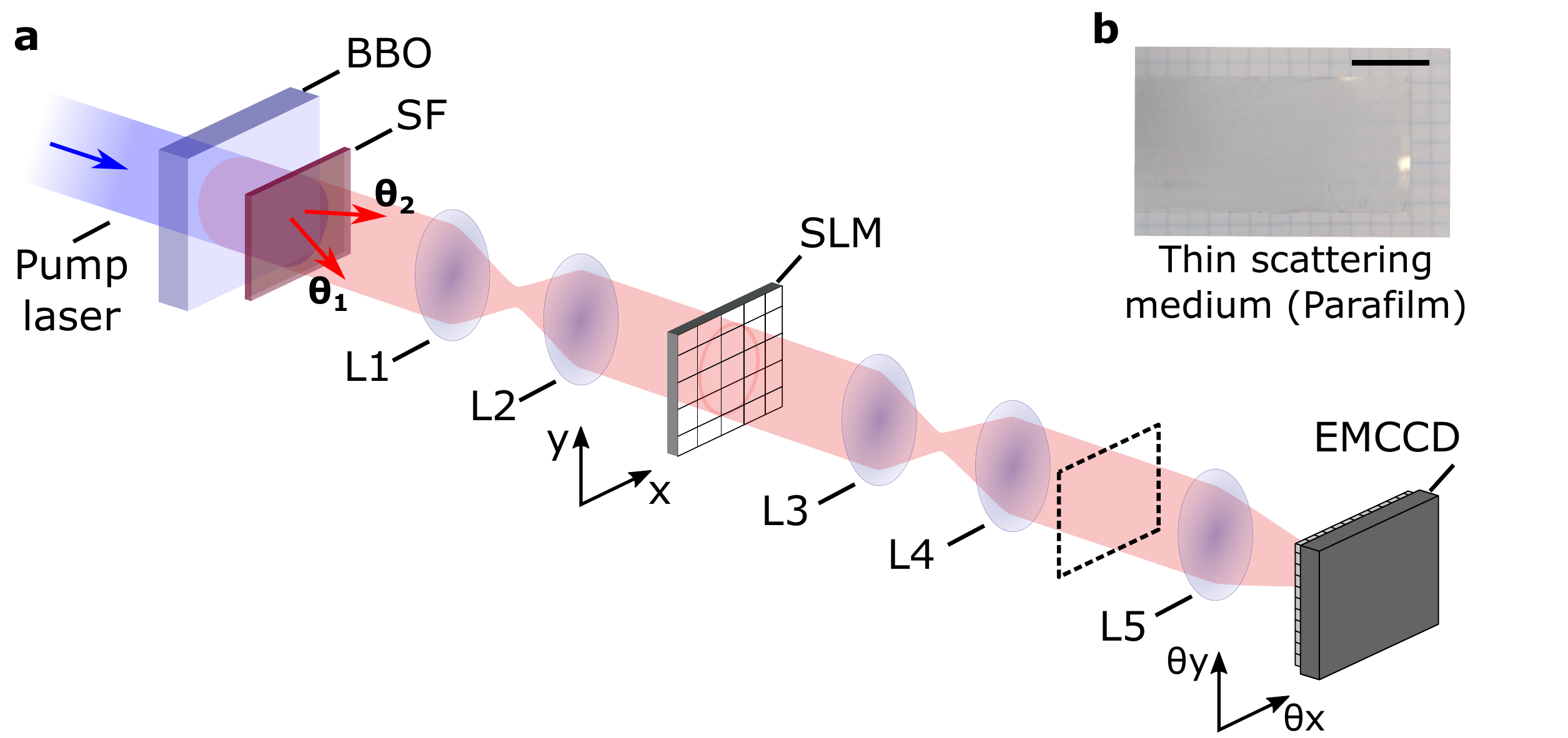}
\caption{\label{Figure1} \textbf{Schematic of the experiment.} \textbf{(a)} Spatially entangled photon pairs are generated by type-I SPDC in a $\beta$-barium borate (BBO) crystal pumped by a collimated continuous-wave laser at $403$ nm. Near-degenerate down-converted photons are selected via spectral filters (SF) at $806 \pm 1.5$ nm. Lenses $L_1$ and $L_2$ image the output surface of the crystal onto the SLM. Pairs of photons are emitted with respective angles denoted $\boldsymbol{\theta_1}$ and $\boldsymbol{\theta_2}$. $L_3$ and $L_4$ image the modulated photons into another optical plane (dashed line), where a scattering medium can be inserted. $L_5$ forms an image of the angular spectrum $\boldsymbol{\theta}=(\theta_x,\theta_y)$ of photon pairs onto an EMCCD camera. The camera enables both direct and correlation intensity measurements. The thin scattering medium  \textbf{(b)} consists of a layer of Parafilm placed on a glass microscope slide. Scale bar is $2$cm. For clarity, the SLM is represented in transmission, while it operates in reflection. }
\end{figure}

\begin{figure*}
\includegraphics[width=0.8 \textwidth]{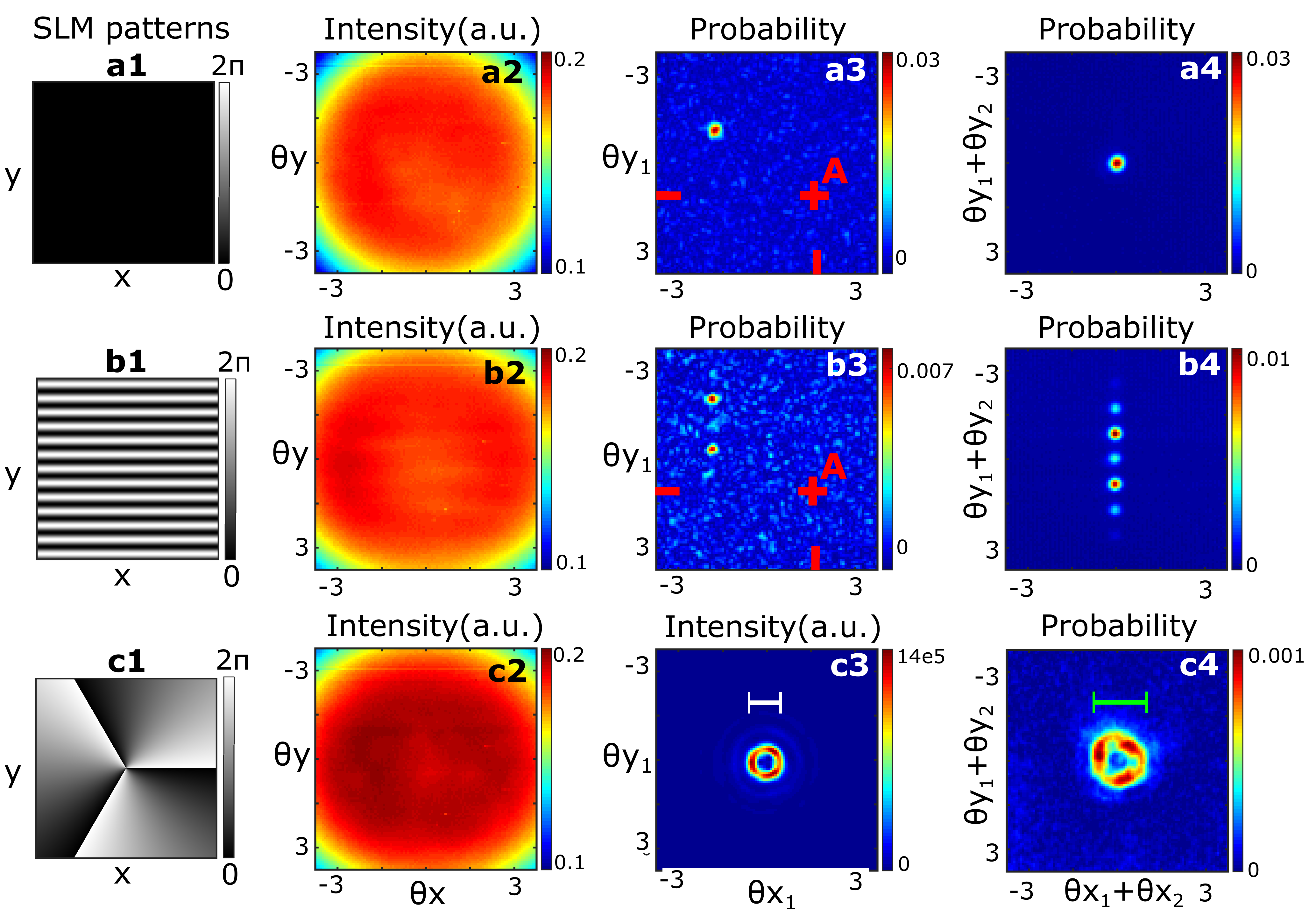}
\caption{\label{Figure2} \textbf{Structuring entanglement by wavefront shaping.} Direct intensity images $I$ (column 2) and joint probability distributions $\Gamma$ (column 3 and 4) are measured under photon-pair illumination using an EMCCD camera~\cite{defienne_general_2018-1,reichert_massively_2018}. Without shaping (\textbf{a1}), anti-correlations in the angular spectrum are visible on a conditional image $\Gamma(\boldsymbol{\theta}_1|\boldsymbol{\theta}_A)$ (\textbf{a3}) [taken for an arbitrarily chosen position $\boldsymbol{\theta}_A = (1.6 \mbox{ mrd},1.1 \mbox{ mrd})$] and on the sum-coordinate projection of $\Gamma$ (\textbf{a4}). A sine phase mask (\textbf{b1}) programmed on the SLM tailors the spatial structure of entanglement into a comb-like pattern, visible on both the conditional image (\textbf{b3}) and on the sum-coordinate projection (\textbf{b4}). A helical SLM phase pattern (\textbf{c1}) produces a ring structure in the sum coordinate projection (\textbf{c4}) with a ring diameter $1.93$mrad (green scale bar). The same experiment performed under classical illumination creates a ring in the direct image (\textbf{c3}) with half the diameter $1.04$mrad (white scale bar). All direct images recorded under quantum illumination (\textbf{a2},\textbf{b2},\textbf{c2}) are independent of the programmed phase patterns. Angular unit is mrad. }
\end{figure*}

In the experiments, we use a phase-only SLM to control the phase of spatially entangled photon pairs and measure its amplitude in the far field (Figure~\ref{Figure1}). In the case of perfectly correlated photons~\cite{abouraddy_entangled-photon_2002}, programming a phase pattern $\phi(\boldsymbol{r})$ tailors the two-photon field $\Psi$ in the reciprocal space as
\begin{equation}
\label{equ1}
\Psi(\boldsymbol{\theta}_1,\boldsymbol{\theta}_2) \propto \iint  e^{2 i \phi(\boldsymbol{r})} e^{-\frac{2 \pi}{\lambda} i \boldsymbol{r} \left[ \boldsymbol{\theta}_1+\boldsymbol{\theta}_2 \right]} d \boldsymbol{r}
\end{equation}
where $\boldsymbol{\theta_1} = \boldsymbol{k_1} \lambda / 2 \pi $ and $\boldsymbol{\theta_2} = \boldsymbol{k_2} \lambda / 2 \pi $ and the angular spectrum (AS) of each photon of a pair, $\boldsymbol{k_1}$ and $\boldsymbol{k_2}$ their respective momentum and $\lambda$ their wavelength (see~\cite{supmat} Section 4). As shown in Figure~\ref{Figure1}.a, the experimental setup is built analogously to a conventional beam shaping system, but the laser is substituted by a quantum source. Spatially entangled photon pairs are generated by type-I spontaneous parametric down conversion (SPDC) in a $\beta$-barium borate (BBO) crystal pumped by a collimated continuous-wave laser at $403$ nm. Near-degenerate down-converted photons are selected via spectral filters (SF) at $\lambda = 806 \pm 1.5$ nm. The output face of the crystal is first imaged onto a phase-only SLM that is itself imaged onto another optical plane (dashed square), where a thin scattering medium will be inserted in the second part of this work. One last lens performs a Fourier transform to map the AS of photons onto the pixels of an electron-multiplied charge-coupled-device (EMCCD) camera. The camera allows (a) direct intensity measurements, providing conventional intensity images $I(\boldsymbol{\theta})$, and (b) correlation intensity measurements, giving the joint probability distribution of photon pairs $\Gamma(\boldsymbol{\theta}_1,\boldsymbol{\theta}_2)=|\Psi(\boldsymbol{\theta}_1,\boldsymbol{\theta}_2)|^2$~\cite{defienne_general_2018-1,reichert_massively_2018}.

\begin{figure*}
\includegraphics[width=0.8 \textwidth]{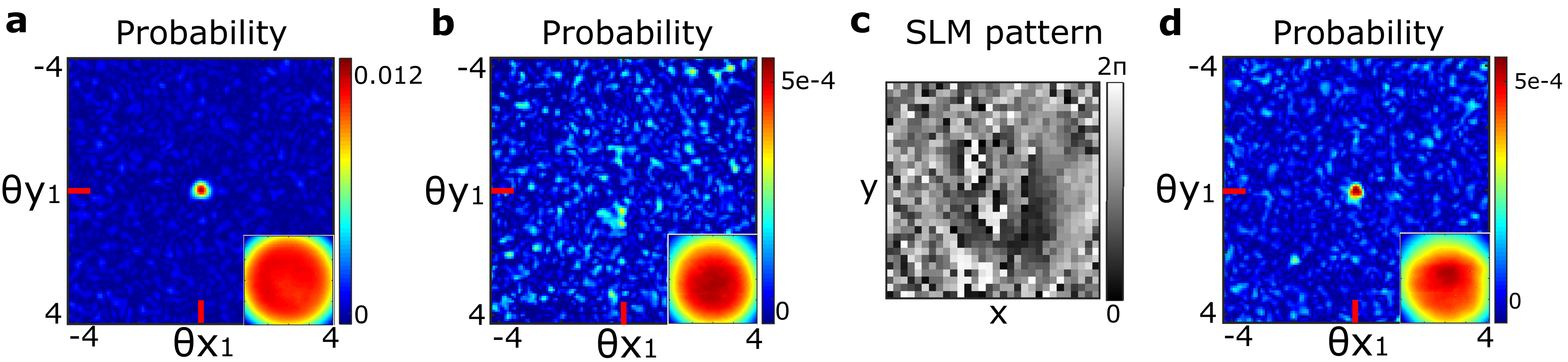}
\caption{\label{Figure3} \textbf{Focusing entanglement through a thin scattering medium}. Conditional image $\Gamma(\boldsymbol{\theta}_1|\boldsymbol{0})$ measured without the medium \textbf{(a)} shows an intense probability peak at $\boldsymbol{\theta}_1=\boldsymbol{0}$. After insertion of the medium, the peak disappears and is replaced by a two-photon speckle pattern \textbf{(b)}. Programming the optimized phase pattern \textbf{(c)}, previously determined using classical coherent light, allows re-focusing of the entanglement at the output of the medium \textbf{(d)}. The shape of the direct intensity image measured at the output (insets) is not affected by the presence of the medium or by the shaping process.}
\end{figure*}

Experimental results are shown in Figure~\ref{Figure2}. When no phase modulation is applied on the SLM, it acts as a mirror, and correlations between photon pairs result only from momentum conservation imposed by the pair generation process (Figure~\ref{Figure2}.a1): when the first photon of a pair is emitted at angle $\boldsymbol{{\theta}}$, its twin is generated at the opposit angle $-\boldsymbol{{\theta}}$. The conditional projection of $\Gamma(\boldsymbol{{\theta_1}}|\boldsymbol{{\theta}_A})$, that represents the probability of detecting one photon when its twin was detected at $\boldsymbol{{\theta_A}}$, shows this anti-correlation property (Figure~\ref{Figure2}.a3). When a sinusoidal phase $\phi(x,y)=\pi/2 \left[ \cos(2 \pi y / \Lambda)+1 \right]$ ($\Lambda = 1.2$mm) is programmed onto the SLM (Figure~\ref{Figure2}.b1), the direct intensity image does not change (Figure~\ref{Figure2}.b2), but the correlation structure between photons does. In this case, $\Gamma(\boldsymbol{{\theta}_1}|\boldsymbol{{\theta_A}})$ takes a comb-like structure centered around $-\boldsymbol{\theta_A}$ with a period proportional to $\lambda/\Lambda$ (Figure~\ref{Figure2}.b3). A clearer representation of the data can be given by using the sum variables $\boldsymbol{{\theta_1}}+\boldsymbol{{\theta_2}}$. This basis provides a better signal-to-noise ratio by integrating $\Gamma(\boldsymbol{{\theta}_1},\boldsymbol{{\theta_2}})$ along the diagonal $(\boldsymbol{{\theta_1}}+\boldsymbol{{\theta_2}})$, as shown in Figure~\ref{Figure2}.b4 (sine phase modulation) and Figure~\ref{Figure2}.a4 (no modulation) [see ~\cite{supmat} section 2 for more details]. In a last example, a helical phase pattern $\phi(x,y)=3 \, \arctan(y/x) $ programmed on the SLM (Figure~\ref{Figure2}.c1) generates a ring on the sum-coordinate projection (Figure~\ref{Figure2}.c4). Interrestingly, the same experiment performed using classical coherent light (Figure~\ref{Figure2}.c3) produces a ring with half the diameter. This factor of two highlights a fundamental difference between classical coherence and quantum coherence of photon pairs, in that the latter accumulate twice the phase during propagation~\cite{brida_experimental_2010}. 

Figure~\ref{Figure2}.c also highlights the fact that knowledge of the classical wavefront can be used to engineer the appropriate quantum structure. This is of enormous benefit for adaptive optics, as the measurement-feedback loop in the quantum case would be practically impossible without it (the Hilbert space is too large, the signal-to-noise ratio too small, etc). Nevertheless, this field is of growing importance, e.g. for sending images and secure information through turbulence~\cite{sit_high-dimensional_2017,bouchard_experimental_2018}. In Figure 3, we show a paradigm example of this problem: re-focusing spatial entanglement that has been randomized by a scattering medium. In this experiment, a layer of parafilm is inserted in an image plane (dashed square) of the setup (Figure~\ref{Figure1}). A comparison between conditional images $\Gamma(\boldsymbol{\theta}_1|\boldsymbol{0})$ taken without the medium (Figure~\ref{Figure3}.a) and after its insertion (Figure~\ref{Figure3}.b) shows a loss of the near-perfect anti-correlations in the AS of photons in favor of a randomly distributed probability pattern called two-photon speckle~\cite{peeters_observation_2010}. To overcome this speckle, we leverage the classical-quantum correspondence in Figure~\ref{Figure2} and the well-established techniques of classical wavefront shaping~\cite{mosk_controlling_2012} to first determine an optimized phase pattern using a coherent source that has the same properties than the photon pairs. This can be done very quickly by interative optimization on the laser intensity at a given pixel of the camera~\cite{vellekoop_focusing_2007} (see~\cite{supmat} Section 5). \textit{The same pattern on the SLM is then used to shape the quantum illumination} (Figure~\ref{Figure3}.c). As shown in Figure~\ref{Figure3}.d, the conditional image $\Gamma(\boldsymbol{\theta}_1|\boldsymbol{\theta}_2=\boldsymbol{0})$ measured at the output shows an intense peak of probability,  demonstrating the re-focusing of pairs in coincidence. As before, quantum coherence is evident in the two-photon field only; the pattern of the direct intensity image measured at the output is affected by neither the medium nor the wavefront shaping process (insets). 

\begin{figure}
\includegraphics[width=1 \columnwidth]{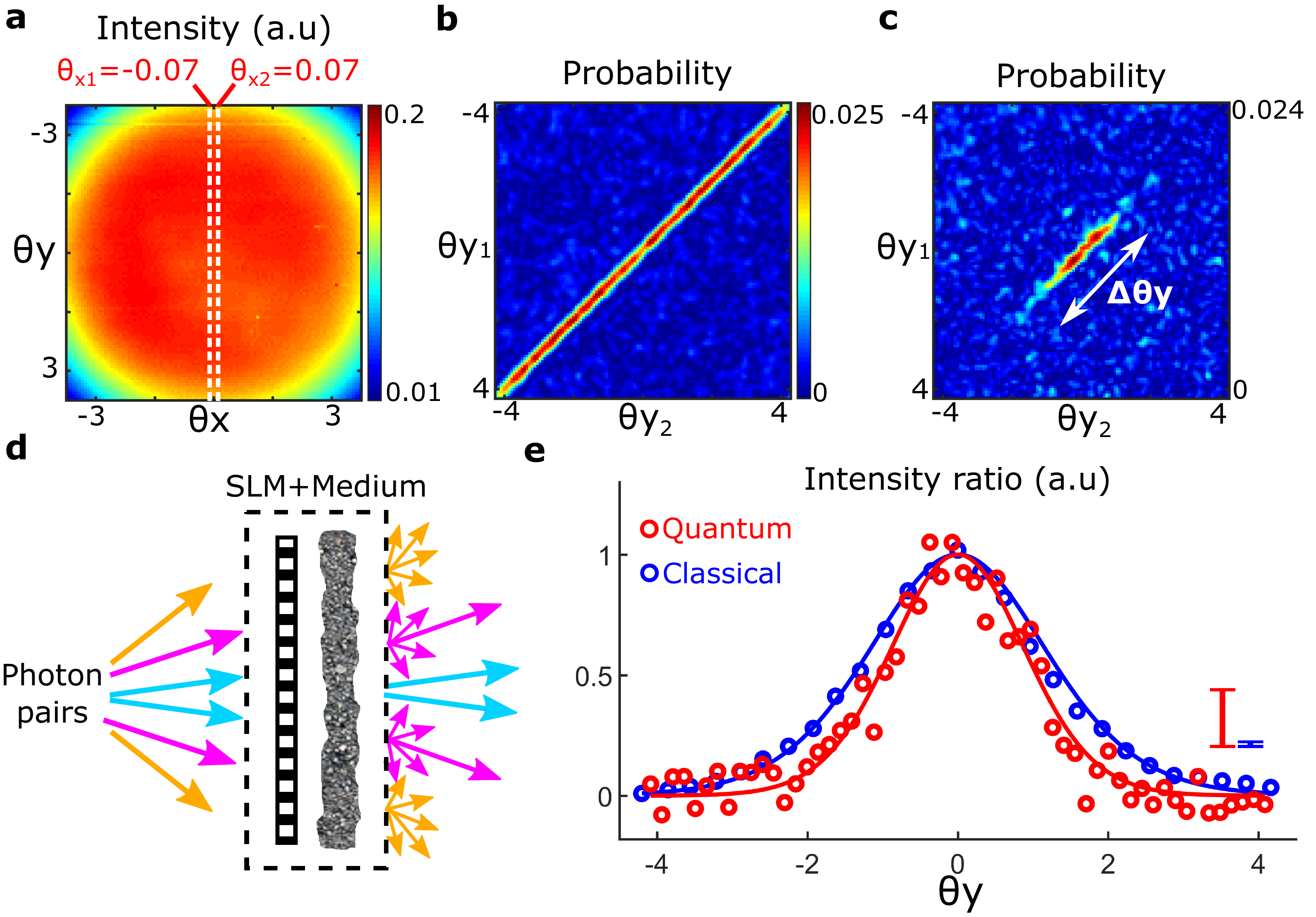}
\caption{\label{Figure4} \textbf{Characterization of the optical memory effect using quantum illumination.} The joint probability distribution of photon pairs is projected onto two columns of pixels located at $\theta_{x_1}=-0.07$ mrd and $\theta_{x_2}=0.07$ mrd on the direct image \textbf{(a)}. Without the scattering medium \textbf{(b)}, $\Gamma(\theta_{y_1},\theta_{y_2}|\theta_{x_1} = -0.07,\theta_{x_2}=0.07)$ shows an intense anti-diagonal, that reflects the anti-correlation in the angular spectrum of photon pairs. With the medium \textbf{(c)}, the optimization technique reconstructs anti-correlations between pairs only in a limited angular range $\Delta \theta_y$, seen as a reduction of the diagonal length. Programming the optimal phase pattern onto the SLM effectively transforms the thin scattering medium into a transparent medium but with a limited field of view \textbf{(d)}. Fitting the focusing ratio $\Gamma(\theta_{x_1}=0,\theta_{y}|\theta_{x_2}=0,-\theta_{y})/\Gamma(\boldsymbol{0}|\boldsymbol{0})$ (red circles) with its corresponding theoretical model (red line) gives $\Delta \theta_y= 1.01 \pm 0.1$ mrd \textbf{(e)}. Classical measurement of the memory effect is represented on the same graph (blue circle) together with its corresponding theoretical model (blue line). For clarity, error bars are shown above the graphs.}
\end{figure}

The optical pattern on the SLM (Figure~\ref{Figure3}.c) is tuned to the scattering medium and gives information about its complexity. Under classical illumination, keeping the pattern the same and changing its incident angle samples a different angular region of the medium. For small angles, the new scattering paths are similar to the old scattering paths, and much of the light is still focused; for larger angles, correlation is lost and a speckle pattern reappears. This range of coherence, known as the optical memory effect~\cite{freund_memory_1988}, provides fundamental insights about the medium (e.g. its scattering mean free path and thickness) and allows imaging though the material~\cite{bertolotti_non-invasive_2012}. Conventionally, the memory effect is characterized with classical light by tilting the compensated wavefront (SLM pattern in Figure~\ref{Figure3}.c) and measuring the falloff in peak intensity  (see~\cite{supmat} section 5). In our work, we show that a measurement of this angular range can be obtained from a single acquisition under quantum illumination. Indeed, at the first-order, spatially entangled photons are incoherent [$g^{(1)}(\boldsymbol{r_1},\boldsymbol{r_2}) = \delta(\boldsymbol{r_1}-\boldsymbol{r_2})$ for perfectly-correlated pairs] and illuminate the medium with a large angular spectrum~\cite{saleh_duality_2000}; Consequently, when structuring quantum light, re-focusing of photon pairs in coincidence occurs not only at the targeted position ($\boldsymbol{\theta}_2=\boldsymbol{0}$) but for larger angles as well, and the memory effect is thus completely characterized by a single measurement of the joint probability distribution. In Figure~\ref{Figure4}, we show the optical memory effect visualized along the $y$-axis (chosen arbitrarily) by projecting the joint probability distribution onto two columns of pixels selected symmetrically from the direct intensity image ($\theta_{x_1} = -0.07$ mrd and $\theta_{x_2} = 0.07$ mrd). As shown in Figure~\ref{Figure4}.c, we observe the presence of a short anti-diagonal at the center of the image, confirming that anti-correlations between pairs are maintained by the wavefront compensation over a finite angular range $\Delta \theta_y$ (in contrast with anti-correlations over the full angular spectrum observed without the medium in Figure~\ref{Figure4}.b). A quantitative analysis performed by fitting the focusing ratio $\Gamma(\theta_{x_1}=0,\theta_{y}|\theta_{x_2}=0,-\theta_{y})/\Gamma(0|0)$ with a theoretical model derived for the quantum case (see~\cite{supmat} Section 6) provides an estimation of the memory effect angle $\Delta \theta_y= 1.01 \pm 0.1$ (Figure~\ref{Figure4}.e - red curve). A comparison with a conventional measurement of the memory effect performed with classical light (blue curve) shows that the nonlocal sampling of the medium by photon pairs (Figure~\ref{Figure4}.d) gives a faster decorrelation. 

Deterministic shaping of entanglement is a promising technique for fundamental physics investigations, such as triggering of coherent processes in mollecules~\cite{roslyak_nonlinear_2009} and plasmons~\cite{altewischer_plasmon-assisted_2002}, and manipulating optical states for quantum storage~\cite{hildner_quantum_2013} and processing~\cite{abouraddy_single-photon_2017}. It provides a straigthforward solution for high-dimensional, entanglement-based quantum communications through imperfect fibers~\cite{amitonova_multimode-fiber-based_2018} and turbulence~\cite{krenn_twisted_2015}. While our experiment involve only spatial entanglement of photon pairs, the methods extend easily to higher orders of quantum coherence~\cite{nagata_beating_2007} and to other degrees of freedom, such as polarization or time~\cite{peer_temporal_2005}. With proper design, adaptive quantum optics can optimize systems where quantum light is beneficial and enable systems where controlled entanglement is required.

\section*{Acknowledgment}
This work was supported by the grants AFOSR FA9550-14-1-0177 and DARPA HR0011-16-C-0027.

\section*{Author contributions}
H.D conceived and performed the experiment; M.R and H.D developed the theory ; all authors analyzed the data and co-wrote the paper.

\nocite{reichert_optimizing_2018,fedorov_gaussian_2009,moreau_realization_2012,tasca_imaging_2012,howell_realization_2004,small_looking_2012,feng_correlations_1988}
\bibliography{scibib}

\begin{thebibliography}{38}%
\makeatletter
\providecommand \@ifxundefined [1]{%
 \@ifx{#1\undefined}
}%
\providecommand \@ifnum [1]{%
 \ifnum #1\expandafter \@firstoftwo
 \else \expandafter \@secondoftwo
 \fi
}%
\providecommand \@ifx [1]{%
 \ifx #1\expandafter \@firstoftwo
 \else \expandafter \@secondoftwo
 \fi
}%
\providecommand \natexlab [1]{#1}%
\providecommand \enquote  [1]{``#1''}%
\providecommand \bibnamefont  [1]{#1}%
\providecommand \bibfnamefont [1]{#1}%
\providecommand \citenamefont [1]{#1}%
\providecommand \href@noop [0]{\@secondoftwo}%
\providecommand \href [0]{\begingroup \@sanitize@url \@href}%
\providecommand \@href[1]{\@@startlink{#1}\@@href}%
\providecommand \@@href[1]{\endgroup#1\@@endlink}%
\providecommand \@sanitize@url [0]{\catcode `\\12\catcode `\$12\catcode
  `\&12\catcode `\#12\catcode `\^12\catcode `\_12\catcode `\%12\relax}%
\providecommand \@@startlink[1]{}%
\providecommand \@@endlink[0]{}%
\providecommand \url  [0]{\begingroup\@sanitize@url \@url }%
\providecommand \@url [1]{\endgroup\@href {#1}{\urlprefix }}%
\providecommand \urlprefix  [0]{URL }%
\providecommand \Eprint [0]{\href }%
\providecommand \doibase [0]{http://dx.doi.org/}%
\providecommand \selectlanguage [0]{\@gobble}%
\providecommand \bibinfo  [0]{\@secondoftwo}%
\providecommand \bibfield  [0]{\@secondoftwo}%
\providecommand \translation [1]{[#1]}%
\providecommand \BibitemOpen [0]{}%
\providecommand \bibitemStop [0]{}%
\providecommand \bibitemNoStop [0]{.\EOS\space}%
\providecommand \EOS [0]{\spacefactor3000\relax}%
\providecommand \BibitemShut  [1]{\csname bibitem#1\endcsname}%
\let\auto@bib@innerbib\@empty
\bibitem [{\citenamefont {Gabor}(1948)}]{gabor_new_1948}%
  \BibitemOpen
  \bibfield  {author} {\bibinfo {author} {\bibfnamefont {D.}~\bibnamefont
  {Gabor}},\ }\href@noop {} {\bibfield  {journal} {\bibinfo  {journal}
  {Nature}\ }\textbf {\bibinfo {volume} {161}},\ \bibinfo {pages} {777}
  (\bibinfo {year} {1948})}\BibitemShut {NoStop}%
\bibitem [{\citenamefont {Zernike}(1955)}]{zernike_how_1955}%
  \BibitemOpen
  \bibfield  {author} {\bibinfo {author} {\bibfnamefont {F.}~\bibnamefont
  {Zernike}},\ }\href {http://www.jstor.org/stable/1682470} {\bibfield
  {journal} {\bibinfo  {journal} {Science}\ }\textbf {\bibinfo {volume}
  {121}},\ \bibinfo {pages} {345} (\bibinfo {year} {1955})}\BibitemShut
  {NoStop}%
\bibitem [{\citenamefont {Yeh}\ and\ \citenamefont
  {Gu}(2009)}]{yeh_optics_2009}%
  \BibitemOpen
  \bibfield  {author} {\bibinfo {author} {\bibfnamefont {P.}~\bibnamefont
  {Yeh}}\ and\ \bibinfo {author} {\bibfnamefont {C.}~\bibnamefont {Gu}},\
  }\href@noop {} {\emph {\bibinfo {title} {Optics of {Liquid} {Crystal}
  {Displays}}}},\ \bibinfo {edition} {2nd}\ ed.\ (\bibinfo  {publisher} {Wiley
  Publishing},\ \bibinfo {year} {2009})\BibitemShut {NoStop}%
\bibitem [{\citenamefont {Weiner}(2000)}]{weiner_femtosecond_2000}%
  \BibitemOpen
  \bibfield  {author} {\bibinfo {author} {\bibfnamefont {A.~M.}\ \bibnamefont
  {Weiner}},\ }\href {\doibase 10.1063/1.1150614} {\bibfield  {journal}
  {\bibinfo  {journal} {Review of Scientific Instruments}\ }\textbf {\bibinfo
  {volume} {71}},\ \bibinfo {pages} {1929} (\bibinfo {year}
  {2000})}\BibitemShut {NoStop}%
\bibitem [{\citenamefont {Klar}\ \emph {et~al.}(2000)\citenamefont {Klar},
  \citenamefont {Jakobs}, \citenamefont {Dyba}, \citenamefont {Egner},\ and\
  \citenamefont {Hell}}]{klar_fluorescence_2000}%
  \BibitemOpen
  \bibfield  {author} {\bibinfo {author} {\bibfnamefont {T.~A.}\ \bibnamefont
  {Klar}}, \bibinfo {author} {\bibfnamefont {S.}~\bibnamefont {Jakobs}},
  \bibinfo {author} {\bibfnamefont {M.}~\bibnamefont {Dyba}}, \bibinfo {author}
  {\bibfnamefont {A.}~\bibnamefont {Egner}}, \ and\ \bibinfo {author}
  {\bibfnamefont {S.~W.}\ \bibnamefont {Hell}},\ }\href {\doibase
  10.1073/pnas.97.15.8206} {\bibfield  {journal} {\bibinfo  {journal}
  {Proceedings of the National Academy of Sciences}\ }\textbf {\bibinfo
  {volume} {97}},\ \bibinfo {pages} {8206} (\bibinfo {year}
  {2000})}\BibitemShut {NoStop}%
\bibitem [{\citenamefont {Leach}\ \emph {et~al.}(2010)\citenamefont {Leach},
  \citenamefont {Jack}, \citenamefont {Romero}, \citenamefont {Jha},
  \citenamefont {Yao}, \citenamefont {Franke-Arnold}, \citenamefont {Ireland},
  \citenamefont {Boyd}, \citenamefont {Barnett},\ and\ \citenamefont
  {Padgett}}]{leach_quantum_2010}%
  \BibitemOpen
  \bibfield  {author} {\bibinfo {author} {\bibfnamefont {J.}~\bibnamefont
  {Leach}}, \bibinfo {author} {\bibfnamefont {B.}~\bibnamefont {Jack}},
  \bibinfo {author} {\bibfnamefont {J.}~\bibnamefont {Romero}}, \bibinfo
  {author} {\bibfnamefont {A.~K.}\ \bibnamefont {Jha}}, \bibinfo {author}
  {\bibfnamefont {A.~M.}\ \bibnamefont {Yao}}, \bibinfo {author} {\bibfnamefont
  {S.}~\bibnamefont {Franke-Arnold}}, \bibinfo {author} {\bibfnamefont {D.~G.}\
  \bibnamefont {Ireland}}, \bibinfo {author} {\bibfnamefont {R.~W.}\
  \bibnamefont {Boyd}}, \bibinfo {author} {\bibfnamefont {S.~M.}\ \bibnamefont
  {Barnett}}, \ and\ \bibinfo {author} {\bibfnamefont {M.~J.}\ \bibnamefont
  {Padgett}},\ }\href {\doibase 10.1126/science.1190523} {\bibfield  {journal}
  {\bibinfo  {journal} {Science}\ }\textbf {\bibinfo {volume} {329}},\ \bibinfo
  {pages} {662} (\bibinfo {year} {2010})}\BibitemShut {NoStop}%
\bibitem [{\citenamefont {Fickler}\ \emph {et~al.}(2012)\citenamefont
  {Fickler}, \citenamefont {Lapkiewicz}, \citenamefont {Plick}, \citenamefont
  {Krenn}, \citenamefont {Schaeff}, \citenamefont {Ramelow},\ and\
  \citenamefont {Zeilinger}}]{fickler_quantum_2012}%
  \BibitemOpen
  \bibfield  {author} {\bibinfo {author} {\bibfnamefont {R.}~\bibnamefont
  {Fickler}}, \bibinfo {author} {\bibfnamefont {R.}~\bibnamefont {Lapkiewicz}},
  \bibinfo {author} {\bibfnamefont {W.~N.}\ \bibnamefont {Plick}}, \bibinfo
  {author} {\bibfnamefont {M.}~\bibnamefont {Krenn}}, \bibinfo {author}
  {\bibfnamefont {C.}~\bibnamefont {Schaeff}}, \bibinfo {author} {\bibfnamefont
  {S.}~\bibnamefont {Ramelow}}, \ and\ \bibinfo {author} {\bibfnamefont
  {A.}~\bibnamefont {Zeilinger}},\ }\href {\doibase 10.1126/science.1227193}
  {\bibfield  {journal} {\bibinfo  {journal} {Science}\ }\textbf {\bibinfo
  {volume} {338}},\ \bibinfo {pages} {640} (\bibinfo {year}
  {2012})}\BibitemShut {NoStop}%
\bibitem [{\citenamefont {Defienne}\ \emph {et~al.}(2016)\citenamefont
  {Defienne}, \citenamefont {Barbieri}, \citenamefont {Walmsley}, \citenamefont
  {Smith},\ and\ \citenamefont {Gigan}}]{defienne_two-photon_2016}%
  \BibitemOpen
  \bibfield  {author} {\bibinfo {author} {\bibfnamefont {H.}~\bibnamefont
  {Defienne}}, \bibinfo {author} {\bibfnamefont {M.}~\bibnamefont {Barbieri}},
  \bibinfo {author} {\bibfnamefont {I.~A.}\ \bibnamefont {Walmsley}}, \bibinfo
  {author} {\bibfnamefont {B.~J.}\ \bibnamefont {Smith}}, \ and\ \bibinfo
  {author} {\bibfnamefont {S.}~\bibnamefont {Gigan}},\ }\href {\doibase
  10.1126/sciadv.1501054} {\bibfield  {journal} {\bibinfo  {journal} {Science
  Advances}\ }\textbf {\bibinfo {volume} {2}},\ \bibinfo {pages} {e1501054}
  (\bibinfo {year} {2016})}\BibitemShut {NoStop}%
\bibitem [{\citenamefont {Wolterink}\ \emph {et~al.}(2016)\citenamefont
  {Wolterink}, \citenamefont {Uppu}, \citenamefont {Ctistis}, \citenamefont
  {Vos}, \citenamefont {Boller},\ and\ \citenamefont
  {Pinkse}}]{wolterink_programmable_2016}%
  \BibitemOpen
  \bibfield  {author} {\bibinfo {author} {\bibfnamefont {T.~A.~W.}\
  \bibnamefont {Wolterink}}, \bibinfo {author} {\bibfnamefont {R.}~\bibnamefont
  {Uppu}}, \bibinfo {author} {\bibfnamefont {G.}~\bibnamefont {Ctistis}},
  \bibinfo {author} {\bibfnamefont {W.~L.}\ \bibnamefont {Vos}}, \bibinfo
  {author} {\bibfnamefont {K.-J.}\ \bibnamefont {Boller}}, \ and\ \bibinfo
  {author} {\bibfnamefont {P.~W.~H.}\ \bibnamefont {Pinkse}},\ }\href {\doibase
  10.1103/PhysRevA.93.053817} {\bibfield  {journal} {\bibinfo  {journal}
  {Physical Review A}\ }\textbf {\bibinfo {volume} {93}},\ \bibinfo {pages}
  {053817} (\bibinfo {year} {2016})}\BibitemShut {NoStop}%
\bibitem [{\citenamefont {Glauber}(1963)}]{glauber_quantum_1963}%
  \BibitemOpen
  \bibfield  {author} {\bibinfo {author} {\bibfnamefont {R.~J.}\ \bibnamefont
  {Glauber}},\ }\href
  {http://journals.aps.org/pr/abstract/10.1103/PhysRev.130.2529} {\bibfield
  {journal} {\bibinfo  {journal} {Physical Review}\ }\textbf {\bibinfo {volume}
  {130}},\ \bibinfo {pages} {2529} (\bibinfo {year} {1963})}\BibitemShut
  {NoStop}%
\bibitem [{\citenamefont {Defienne}\ \emph {et~al.}(2018)\citenamefont
  {Defienne}, \citenamefont {Reichert},\ and\ \citenamefont
  {Fleischer}}]{defienne_general_2018-1}%
  \BibitemOpen
  \bibfield  {author} {\bibinfo {author} {\bibfnamefont {H.}~\bibnamefont
  {Defienne}}, \bibinfo {author} {\bibfnamefont {M.}~\bibnamefont {Reichert}},
  \ and\ \bibinfo {author} {\bibfnamefont {J.~W.}\ \bibnamefont {Fleischer}},\
  }\href@noop {} {\bibfield  {journal} {\bibinfo  {journal} {Physical review
  letters}\ }\textbf {\bibinfo {volume} {120}},\ \bibinfo {pages} {203604}
  (\bibinfo {year} {2018})}\BibitemShut {NoStop}%
\bibitem [{\citenamefont {Reichert}\ \emph
  {et~al.}(2018{\natexlab{a}})\citenamefont {Reichert}, \citenamefont
  {Defienne},\ and\ \citenamefont {Fleischer}}]{reichert_massively_2018}%
  \BibitemOpen
  \bibfield  {author} {\bibinfo {author} {\bibfnamefont {M.}~\bibnamefont
  {Reichert}}, \bibinfo {author} {\bibfnamefont {H.}~\bibnamefont {Defienne}},
  \ and\ \bibinfo {author} {\bibfnamefont {J.~W.}\ \bibnamefont {Fleischer}},\
  }\href@noop {} {\bibfield  {journal} {\bibinfo  {journal} {Scientific
  reports}\ }\textbf {\bibinfo {volume} {8}},\ \bibinfo {pages} {7925}
  (\bibinfo {year} {2018}{\natexlab{a}})}\BibitemShut {NoStop}%
\bibitem [{\citenamefont {Abouraddy}\ \emph {et~al.}(2002)\citenamefont
  {Abouraddy}, \citenamefont {Saleh}, \citenamefont {Sergienko},\ and\
  \citenamefont {Teich}}]{abouraddy_entangled-photon_2002}%
  \BibitemOpen
  \bibfield  {author} {\bibinfo {author} {\bibfnamefont {A.~F.}\ \bibnamefont
  {Abouraddy}}, \bibinfo {author} {\bibfnamefont {B.~E.~A.}\ \bibnamefont
  {Saleh}}, \bibinfo {author} {\bibfnamefont {A.~V.}\ \bibnamefont
  {Sergienko}}, \ and\ \bibinfo {author} {\bibfnamefont {M.~C.}\ \bibnamefont
  {Teich}},\ }\href {\doibase 10.1364/JOSAB.19.001174} {\bibfield  {journal}
  {\bibinfo  {journal} {JOSA B}\ }\textbf {\bibinfo {volume} {19}},\ \bibinfo
  {pages} {1174} (\bibinfo {year} {2002})}\BibitemShut {NoStop}%
\bibitem [{sup()}]{supmat}%
  \BibitemOpen
  \href@noop {} {\bibinfo  {journal} {See Supplemental Material at
  http://link.aps.org/ supplemental/10.1103/PhysRevLett.121.233601 for in-depth
  methods, additional results and theoretical demonstrations, which includes
  Refs. [32-38].}\ }\BibitemShut {NoStop}%
\bibitem [{\citenamefont {Brida}\ \emph {et~al.}(2010)\citenamefont {Brida},
  \citenamefont {Genovese},\ and\ \citenamefont
  {Berchera}}]{brida_experimental_2010}%
  \BibitemOpen
\bibfield  {journal} {  }\bibfield  {author} {\bibinfo {author} {\bibfnamefont
  {G.}~\bibnamefont {Brida}}, \bibinfo {author} {\bibfnamefont
  {M.}~\bibnamefont {Genovese}}, \ and\ \bibinfo {author} {\bibfnamefont
  {I.~R.}\ \bibnamefont {Berchera}},\ }\href@noop {} {\bibfield  {journal}
  {\bibinfo  {journal} {Nature Photonics}\ }\textbf {\bibinfo {volume} {4}},\
  \bibinfo {pages} {227} (\bibinfo {year} {2010})}\BibitemShut {NoStop}%
\bibitem [{\citenamefont {Sit}\ \emph {et~al.}(2017)\citenamefont {Sit},
  \citenamefont {Bouchard}, \citenamefont {Fickler}, \citenamefont
  {Gagnon-Bischoff}, \citenamefont {Larocque}, \citenamefont {Heshami},
  \citenamefont {Elser}, \citenamefont {Peuntinger}, \citenamefont {Günthner},
  \citenamefont {Heim}, \citenamefont {Marquardt}, \citenamefont {Leuchs},
  \citenamefont {Boyd},\ and\ \citenamefont
  {Karimi}}]{sit_high-dimensional_2017}%
  \BibitemOpen
  \bibfield  {author} {\bibinfo {author} {\bibfnamefont {A.}~\bibnamefont
  {Sit}}, \bibinfo {author} {\bibfnamefont {F.}~\bibnamefont {Bouchard}},
  \bibinfo {author} {\bibfnamefont {R.}~\bibnamefont {Fickler}}, \bibinfo
  {author} {\bibfnamefont {J.}~\bibnamefont {Gagnon-Bischoff}}, \bibinfo
  {author} {\bibfnamefont {H.}~\bibnamefont {Larocque}}, \bibinfo {author}
  {\bibfnamefont {K.}~\bibnamefont {Heshami}}, \bibinfo {author} {\bibfnamefont
  {D.}~\bibnamefont {Elser}}, \bibinfo {author} {\bibfnamefont
  {C.}~\bibnamefont {Peuntinger}}, \bibinfo {author} {\bibfnamefont
  {K.}~\bibnamefont {Günthner}}, \bibinfo {author} {\bibfnamefont
  {B.}~\bibnamefont {Heim}}, \bibinfo {author} {\bibfnamefont {C.}~\bibnamefont
  {Marquardt}}, \bibinfo {author} {\bibfnamefont {G.}~\bibnamefont {Leuchs}},
  \bibinfo {author} {\bibfnamefont {R.~W.}\ \bibnamefont {Boyd}}, \ and\
  \bibinfo {author} {\bibfnamefont {E.}~\bibnamefont {Karimi}},\ }\href
  {\doibase 10.1364/OPTICA.4.001006} {\bibfield  {journal} {\bibinfo  {journal}
  {Optica}\ }\textbf {\bibinfo {volume} {4}},\ \bibinfo {pages} {1006}
  (\bibinfo {year} {2017})}\BibitemShut {NoStop}%
\bibitem [{\citenamefont {Bouchard}\ \emph {et~al.}(2018)\citenamefont
  {Bouchard}, \citenamefont {Heshami}, \citenamefont {England}, \citenamefont
  {Fickler}, \citenamefont {Boyd}, \citenamefont {Englert}, \citenamefont
  {Sánchez-Soto},\ and\ \citenamefont {Karimi}}]{bouchard_experimental_2018}%
  \BibitemOpen
  \bibfield  {author} {\bibinfo {author} {\bibfnamefont {F.}~\bibnamefont
  {Bouchard}}, \bibinfo {author} {\bibfnamefont {K.}~\bibnamefont {Heshami}},
  \bibinfo {author} {\bibfnamefont {D.}~\bibnamefont {England}}, \bibinfo
  {author} {\bibfnamefont {R.}~\bibnamefont {Fickler}}, \bibinfo {author}
  {\bibfnamefont {R.~W.}\ \bibnamefont {Boyd}}, \bibinfo {author}
  {\bibfnamefont {B.-G.}\ \bibnamefont {Englert}}, \bibinfo {author}
  {\bibfnamefont {L.~L.}\ \bibnamefont {Sánchez-Soto}}, \ and\ \bibinfo
  {author} {\bibfnamefont {E.}~\bibnamefont {Karimi}},\ }\href
  {http://arxiv.org/abs/1802.05773} {\bibfield  {journal} {\bibinfo  {journal}
  {arXiv:1802.05773 [quant-ph]}\ } (\bibinfo {year} {2018})},\ \bibinfo {note}
  {arXiv: 1802.05773}\BibitemShut {NoStop}%
\bibitem [{\citenamefont {Peeters}\ \emph {et~al.}(2010)\citenamefont
  {Peeters}, \citenamefont {Moerman},\ and\ \citenamefont {van
  Exter}}]{peeters_observation_2010}%
  \BibitemOpen
  \bibfield  {author} {\bibinfo {author} {\bibfnamefont {W.~H.}\ \bibnamefont
  {Peeters}}, \bibinfo {author} {\bibfnamefont {J.~J.~D.}\ \bibnamefont
  {Moerman}}, \ and\ \bibinfo {author} {\bibfnamefont {M.~P.}\ \bibnamefont
  {van Exter}},\ }\href {\doibase 10.1103/PhysRevLett.104.173601} {\bibfield
  {journal} {\bibinfo  {journal} {Physical Review Letters}\ }\textbf {\bibinfo
  {volume} {104}} (\bibinfo {year} {2010}),\
  10.1103/PhysRevLett.104.173601}\BibitemShut {NoStop}%
\bibitem [{\citenamefont {Mosk}\ \emph {et~al.}(2012)\citenamefont {Mosk},
  \citenamefont {Lagendijk}, \citenamefont {Lerosey},\ and\ \citenamefont
  {Fink}}]{mosk_controlling_2012}%
  \BibitemOpen
  \bibfield  {author} {\bibinfo {author} {\bibfnamefont {A.~P.}\ \bibnamefont
  {Mosk}}, \bibinfo {author} {\bibfnamefont {A.}~\bibnamefont {Lagendijk}},
  \bibinfo {author} {\bibfnamefont {G.}~\bibnamefont {Lerosey}}, \ and\
  \bibinfo {author} {\bibfnamefont {M.}~\bibnamefont {Fink}},\ }\href {\doibase
  10.1038/nphoton.2012.88} {\bibfield  {journal} {\bibinfo  {journal} {Nature
  Photonics}\ }\textbf {\bibinfo {volume} {6}},\ \bibinfo {pages} {283}
  (\bibinfo {year} {2012})}\BibitemShut {NoStop}%
\bibitem [{\citenamefont {Vellekoop}\ and\ \citenamefont
  {Mosk}(2007)}]{vellekoop_focusing_2007}%
  \BibitemOpen
  \bibfield  {author} {\bibinfo {author} {\bibfnamefont {I.~M.}\ \bibnamefont
  {Vellekoop}}\ and\ \bibinfo {author} {\bibfnamefont {A.~P.}\ \bibnamefont
  {Mosk}},\ }\href {\doibase 10.1364/OL.32.002309} {\bibfield  {journal}
  {\bibinfo  {journal} {Optics Letters}\ }\textbf {\bibinfo {volume} {32}},\
  \bibinfo {pages} {2309} (\bibinfo {year} {2007})}\BibitemShut {NoStop}%
\bibitem [{\citenamefont {Freund}\ \emph {et~al.}(1988)\citenamefont {Freund},
  \citenamefont {Rosenbluh},\ and\ \citenamefont {Feng}}]{freund_memory_1988}%
  \BibitemOpen
  \bibfield  {author} {\bibinfo {author} {\bibfnamefont {I.}~\bibnamefont
  {Freund}}, \bibinfo {author} {\bibfnamefont {M.}~\bibnamefont {Rosenbluh}}, \
  and\ \bibinfo {author} {\bibfnamefont {S.}~\bibnamefont {Feng}},\ }\href
  {\doibase 10.1103/PhysRevLett.61.2328} {\bibfield  {journal} {\bibinfo
  {journal} {Physical Review Letters}\ }\textbf {\bibinfo {volume} {61}},\
  \bibinfo {pages} {2328} (\bibinfo {year} {1988})}\BibitemShut {NoStop}%
\bibitem [{\citenamefont {Bertolotti}\ \emph {et~al.}(2012)\citenamefont
  {Bertolotti}, \citenamefont {van Putten}, \citenamefont {Blum}, \citenamefont
  {Lagendijk}, \citenamefont {Vos},\ and\ \citenamefont
  {Mosk}}]{bertolotti_non-invasive_2012}%
  \BibitemOpen
  \bibfield  {author} {\bibinfo {author} {\bibfnamefont {J.}~\bibnamefont
  {Bertolotti}}, \bibinfo {author} {\bibfnamefont {E.~G.}\ \bibnamefont {van
  Putten}}, \bibinfo {author} {\bibfnamefont {C.}~\bibnamefont {Blum}},
  \bibinfo {author} {\bibfnamefont {A.}~\bibnamefont {Lagendijk}}, \bibinfo
  {author} {\bibfnamefont {W.~L.}\ \bibnamefont {Vos}}, \ and\ \bibinfo
  {author} {\bibfnamefont {A.~P.}\ \bibnamefont {Mosk}},\ }\href@noop {}
  {\bibfield  {journal} {\bibinfo  {journal} {Nature}\ }\textbf {\bibinfo
  {volume} {491}},\ \bibinfo {pages} {232} (\bibinfo {year}
  {2012})}\BibitemShut {NoStop}%
\bibitem [{\citenamefont {Saleh}\ \emph {et~al.}(2000)\citenamefont {Saleh},
  \citenamefont {Abouraddy}, \citenamefont {Sergienko},\ and\ \citenamefont
  {Teich}}]{saleh_duality_2000}%
  \BibitemOpen
  \bibfield  {author} {\bibinfo {author} {\bibfnamefont {B.~E.~A.}\
  \bibnamefont {Saleh}}, \bibinfo {author} {\bibfnamefont {A.~F.}\ \bibnamefont
  {Abouraddy}}, \bibinfo {author} {\bibfnamefont {A.~V.}\ \bibnamefont
  {Sergienko}}, \ and\ \bibinfo {author} {\bibfnamefont {M.~C.}\ \bibnamefont
  {Teich}},\ }\href {\doibase 10.1103/PhysRevA.62.043816} {\bibfield  {journal}
  {\bibinfo  {journal} {Physical Review A}\ }\textbf {\bibinfo {volume} {62}},\
  \bibinfo {pages} {043816} (\bibinfo {year} {2000})}\BibitemShut {NoStop}%
\bibitem [{\citenamefont {Roslyak}\ \emph {et~al.}(2009)\citenamefont
  {Roslyak}, \citenamefont {Marx},\ and\ \citenamefont
  {Mukame}}]{roslyak_nonlinear_2009}%
  \BibitemOpen
  \bibfield  {author} {\bibinfo {author} {\bibfnamefont {O.}~\bibnamefont
  {Roslyak}}, \bibinfo {author} {\bibfnamefont {C.~A.}\ \bibnamefont {Marx}}, \
  and\ \bibinfo {author} {\bibfnamefont {S.}~\bibnamefont {Mukame}},\ }\href
  {\doibase 10.1103/PhysRevA.79.033832} {\bibfield  {journal} {\bibinfo
  {journal} {Physical review. A}\ }\textbf {\bibinfo {volume} {79}},\ \bibinfo
  {pages} {33832} (\bibinfo {year} {2009})}\BibitemShut {NoStop}%
\bibitem [{\citenamefont {Altewischer}\ \emph {et~al.}(2002)\citenamefont
  {Altewischer}, \citenamefont {van Exter},\ and\ \citenamefont
  {Woerdman}}]{altewischer_plasmon-assisted_2002}%
  \BibitemOpen
  \bibfield  {author} {\bibinfo {author} {\bibfnamefont {E.}~\bibnamefont
  {Altewischer}}, \bibinfo {author} {\bibfnamefont {M.~P.}\ \bibnamefont {van
  Exter}}, \ and\ \bibinfo {author} {\bibfnamefont {J.~P.}\ \bibnamefont
  {Woerdman}},\ }\href {\doibase 10.1038/nature00869} {\bibfield  {journal}
  {\bibinfo  {journal} {Nature}\ }\textbf {\bibinfo {volume} {418}},\ \bibinfo
  {pages} {304} (\bibinfo {year} {2002})}\BibitemShut {NoStop}%
\bibitem [{\citenamefont {Hildner}\ \emph {et~al.}(2013)\citenamefont
  {Hildner}, \citenamefont {Brinks}, \citenamefont {Nieder}, \citenamefont
  {Cogdell},\ and\ \citenamefont {Hulst}}]{hildner_quantum_2013}%
  \BibitemOpen
  \bibfield  {author} {\bibinfo {author} {\bibfnamefont {R.}~\bibnamefont
  {Hildner}}, \bibinfo {author} {\bibfnamefont {D.}~\bibnamefont {Brinks}},
  \bibinfo {author} {\bibfnamefont {J.~B.}\ \bibnamefont {Nieder}}, \bibinfo
  {author} {\bibfnamefont {R.~J.}\ \bibnamefont {Cogdell}}, \ and\ \bibinfo
  {author} {\bibfnamefont {N.~F.~v.}\ \bibnamefont {Hulst}},\ }\href {\doibase
  10.1126/science.1235820} {\bibfield  {journal} {\bibinfo  {journal}
  {Science}\ }\textbf {\bibinfo {volume} {340}},\ \bibinfo {pages} {1448}
  (\bibinfo {year} {2013})}\BibitemShut {NoStop}%
\bibitem [{\citenamefont {Abouraddy}\ \emph {et~al.}(2017)\citenamefont
  {Abouraddy}, \citenamefont {Saleh}, \citenamefont {Giuseppe},\ and\
  \citenamefont {Kagalwala}}]{abouraddy_single-photon_2017}%
  \BibitemOpen
  \bibfield  {author} {\bibinfo {author} {\bibfnamefont {A.~F.}\ \bibnamefont
  {Abouraddy}}, \bibinfo {author} {\bibfnamefont {B.~E.~A.}\ \bibnamefont
  {Saleh}}, \bibinfo {author} {\bibfnamefont {G.}~\bibnamefont {Giuseppe}}, \
  and\ \bibinfo {author} {\bibfnamefont {K.~H.}\ \bibnamefont {Kagalwala}},\
  }\href {\doibase 10.1038/s41467-017-00580-x} {\bibfield  {journal} {\bibinfo
  {journal} {Nature Communications}\ }\textbf {\bibinfo {volume} {8}},\
  \bibinfo {pages} {739} (\bibinfo {year} {2017})}\BibitemShut {NoStop}%
\bibitem [{\citenamefont {Amitonova}\ \emph {et~al.}(2018)\citenamefont
  {Amitonova}, \citenamefont {Tentrup}, \citenamefont {Vellekoop},\ and\
  \citenamefont {Pinkse}}]{amitonova_multimode-fiber-based_2018}%
  \BibitemOpen
  \bibfield  {author} {\bibinfo {author} {\bibfnamefont {L.~V.}\ \bibnamefont
  {Amitonova}}, \bibinfo {author} {\bibfnamefont {T.~B.~H.}\ \bibnamefont
  {Tentrup}}, \bibinfo {author} {\bibfnamefont {I.~M.}\ \bibnamefont
  {Vellekoop}}, \ and\ \bibinfo {author} {\bibfnamefont {P.~W.~H.}\
  \bibnamefont {Pinkse}},\ }\href {http://arxiv.org/abs/1801.07180} {\bibfield
  {journal} {\bibinfo  {journal} {arXiv:1801.07180 [physics,
  physics:quant-ph]}\ } (\bibinfo {year} {2018})},\ \bibinfo {note} {arXiv:
  1801.07180}\BibitemShut {NoStop}%
\bibitem [{\citenamefont {Krenn}\ \emph {et~al.}(2015)\citenamefont {Krenn},
  \citenamefont {Handsteiner}, \citenamefont {Fink}, \citenamefont {Fickler},\
  and\ \citenamefont {Zeilinger}}]{krenn_twisted_2015}%
  \BibitemOpen
  \bibfield  {author} {\bibinfo {author} {\bibfnamefont {M.}~\bibnamefont
  {Krenn}}, \bibinfo {author} {\bibfnamefont {J.}~\bibnamefont {Handsteiner}},
  \bibinfo {author} {\bibfnamefont {M.}~\bibnamefont {Fink}}, \bibinfo {author}
  {\bibfnamefont {R.}~\bibnamefont {Fickler}}, \ and\ \bibinfo {author}
  {\bibfnamefont {A.}~\bibnamefont {Zeilinger}},\ }\href {\doibase
  10.1073/pnas.1517574112} {\bibfield  {journal} {\bibinfo  {journal}
  {Proceedings of the National Academy of Sciences}\ }\textbf {\bibinfo
  {volume} {112}},\ \bibinfo {pages} {14197} (\bibinfo {year}
  {2015})}\BibitemShut {NoStop}%
\bibitem [{\citenamefont {Nagata}\ \emph {et~al.}(2007)\citenamefont {Nagata},
  \citenamefont {Okamoto}, \citenamefont {O'Brien}, \citenamefont {Sasaki},\
  and\ \citenamefont {Takeuchi}}]{nagata_beating_2007}%
  \BibitemOpen
  \bibfield  {author} {\bibinfo {author} {\bibfnamefont {T.}~\bibnamefont
  {Nagata}}, \bibinfo {author} {\bibfnamefont {R.}~\bibnamefont {Okamoto}},
  \bibinfo {author} {\bibfnamefont {J.~L.}\ \bibnamefont {O'Brien}}, \bibinfo
  {author} {\bibfnamefont {K.}~\bibnamefont {Sasaki}}, \ and\ \bibinfo {author}
  {\bibfnamefont {S.}~\bibnamefont {Takeuchi}},\ }\href {\doibase
  10.1126/science.1138007} {\bibfield  {journal} {\bibinfo  {journal}
  {Science}\ }\textbf {\bibinfo {volume} {316}},\ \bibinfo {pages} {726}
  (\bibinfo {year} {2007})}\BibitemShut {NoStop}%
\bibitem [{\citenamefont {Pe'Er}\ \emph {et~al.}(2005)\citenamefont {Pe'Er},
  \citenamefont {Dayan}, \citenamefont {Friesem},\ and\ \citenamefont
  {Silberberg}}]{peer_temporal_2005}%
  \BibitemOpen
  \bibfield  {author} {\bibinfo {author} {\bibfnamefont {A.}~\bibnamefont
  {Pe'Er}}, \bibinfo {author} {\bibfnamefont {B.}~\bibnamefont {Dayan}},
  \bibinfo {author} {\bibfnamefont {A.~A.}\ \bibnamefont {Friesem}}, \ and\
  \bibinfo {author} {\bibfnamefont {Y.}~\bibnamefont {Silberberg}},\
  }\href@noop {} {\bibfield  {journal} {\bibinfo  {journal} {Physical review
  letters}\ }\textbf {\bibinfo {volume} {94}},\ \bibinfo {pages} {073601}
  (\bibinfo {year} {2005})}\BibitemShut {NoStop}%
\bibitem [{\citenamefont {Reichert}\ \emph
  {et~al.}(2018{\natexlab{b}})\citenamefont {Reichert}, \citenamefont
  {Defienne},\ and\ \citenamefont {Fleischer}}]{reichert_optimizing_2018}%
  \BibitemOpen
  \bibfield  {author} {\bibinfo {author} {\bibfnamefont {M.}~\bibnamefont
  {Reichert}}, \bibinfo {author} {\bibfnamefont {H.}~\bibnamefont {Defienne}},
  \ and\ \bibinfo {author} {\bibfnamefont {J.~W.}\ \bibnamefont {Fleischer}},\
  }\href {\doibase 10.1103/PhysRevA.98.013841} {\bibfield  {journal} {\bibinfo
  {journal} {Physical Review A}\ }\textbf {\bibinfo {volume} {98}},\ \bibinfo
  {pages} {013841} (\bibinfo {year} {2018}{\natexlab{b}})}\BibitemShut
  {NoStop}%
\bibitem [{\citenamefont {Fedorov}\ \emph {et~al.}(2009)\citenamefont
  {Fedorov}, \citenamefont {Mikhailova},\ and\ \citenamefont
  {Volkov}}]{fedorov_gaussian_2009}%
  \BibitemOpen
  \bibfield  {author} {\bibinfo {author} {\bibfnamefont {M.~V.}\ \bibnamefont
  {Fedorov}}, \bibinfo {author} {\bibfnamefont {Y.~M.}\ \bibnamefont
  {Mikhailova}}, \ and\ \bibinfo {author} {\bibfnamefont {P.~A.}\ \bibnamefont
  {Volkov}},\ }\href {\doibase 10.1088/0953-4075/42/17/175503} {\bibfield
  {journal} {\bibinfo  {journal} {Journal of Physics B: Atomic, Molecular and
  Optical Physics}\ }\textbf {\bibinfo {volume} {42}},\ \bibinfo {pages}
  {175503} (\bibinfo {year} {2009})}\BibitemShut {NoStop}%
\bibitem [{\citenamefont {Moreau}\ \emph {et~al.}(2012)\citenamefont {Moreau},
  \citenamefont {Mougin-Sisini}, \citenamefont {Devaux},\ and\ \citenamefont
  {Lantz}}]{moreau_realization_2012}%
  \BibitemOpen
  \bibfield  {author} {\bibinfo {author} {\bibfnamefont {P.-A.}\ \bibnamefont
  {Moreau}}, \bibinfo {author} {\bibfnamefont {J.}~\bibnamefont
  {Mougin-Sisini}}, \bibinfo {author} {\bibfnamefont {F.}~\bibnamefont
  {Devaux}}, \ and\ \bibinfo {author} {\bibfnamefont {E.}~\bibnamefont
  {Lantz}},\ }\href {\doibase 10.1103/PhysRevA.86.010101} {\bibfield  {journal}
  {\bibinfo  {journal} {Physical Review A}\ }\textbf {\bibinfo {volume} {86}},\
  \bibinfo {pages} {010101} (\bibinfo {year} {2012})}\BibitemShut {NoStop}%
\bibitem [{\citenamefont {Tasca}\ \emph {et~al.}(2012)\citenamefont {Tasca},
  \citenamefont {Izdebski}, \citenamefont {Buller}, \citenamefont {Leach},
  \citenamefont {Agnew}, \citenamefont {Padgett}, \citenamefont {Edgar},
  \citenamefont {Warburton},\ and\ \citenamefont {Boyd}}]{tasca_imaging_2012}%
  \BibitemOpen
  \bibfield  {author} {\bibinfo {author} {\bibfnamefont {D.~S.}\ \bibnamefont
  {Tasca}}, \bibinfo {author} {\bibfnamefont {F.}~\bibnamefont {Izdebski}},
  \bibinfo {author} {\bibfnamefont {G.~S.}\ \bibnamefont {Buller}}, \bibinfo
  {author} {\bibfnamefont {J.}~\bibnamefont {Leach}}, \bibinfo {author}
  {\bibfnamefont {M.}~\bibnamefont {Agnew}}, \bibinfo {author} {\bibfnamefont
  {M.~J.}\ \bibnamefont {Padgett}}, \bibinfo {author} {\bibfnamefont {M.~P.}\
  \bibnamefont {Edgar}}, \bibinfo {author} {\bibfnamefont {R.~E.}\ \bibnamefont
  {Warburton}}, \ and\ \bibinfo {author} {\bibfnamefont {R.~W.}\ \bibnamefont
  {Boyd}},\ }\href {\doibase 10.1038/ncomms1988} {\bibfield  {journal}
  {\bibinfo  {journal} {Nature Communications}\ }\textbf {\bibinfo {volume}
  {3}},\ \bibinfo {pages} {984} (\bibinfo {year} {2012})}\BibitemShut {NoStop}%
\bibitem [{\citenamefont {Howell}\ \emph {et~al.}(2004)\citenamefont {Howell},
  \citenamefont {Bennink}, \citenamefont {Bentley},\ and\ \citenamefont
  {Boyd}}]{howell_realization_2004}%
  \BibitemOpen
  \bibfield  {author} {\bibinfo {author} {\bibfnamefont {J.~C.}\ \bibnamefont
  {Howell}}, \bibinfo {author} {\bibfnamefont {R.~S.}\ \bibnamefont {Bennink}},
  \bibinfo {author} {\bibfnamefont {S.~J.}\ \bibnamefont {Bentley}}, \ and\
  \bibinfo {author} {\bibfnamefont {R.~W.}\ \bibnamefont {Boyd}},\ }\href
  {\doibase 10.1103/PhysRevLett.92.210403} {\bibfield  {journal} {\bibinfo
  {journal} {Physical Review Letters}\ }\textbf {\bibinfo {volume} {92}},\
  \bibinfo {pages} {210403} (\bibinfo {year} {2004})}\BibitemShut {NoStop}%
\bibitem [{\citenamefont {Small}\ \emph {et~al.}(2012)\citenamefont {Small},
  \citenamefont {Katz},\ and\ \citenamefont {Silberberg}}]{small_looking_2012}%
  \BibitemOpen
  \bibfield  {author} {\bibinfo {author} {\bibfnamefont {E.}~\bibnamefont
  {Small}}, \bibinfo {author} {\bibfnamefont {O.}~\bibnamefont {Katz}}, \ and\
  \bibinfo {author} {\bibfnamefont {Y.}~\bibnamefont {Silberberg}},\ }\href
  {\doibase 10.1038/nphoton.2012.150} {\bibfield  {journal} {\bibinfo
  {journal} {Nature Photonics}\ }\textbf {\bibinfo {volume} {6}},\ \bibinfo
  {pages} {549} (\bibinfo {year} {2012})}\BibitemShut {NoStop}%
\bibitem [{\citenamefont {Feng}\ \emph {et~al.}(1988)\citenamefont {Feng},
  \citenamefont {Kane}, \citenamefont {Lee},\ and\ \citenamefont
  {Stone}}]{feng_correlations_1988}%
  \BibitemOpen
  \bibfield  {author} {\bibinfo {author} {\bibfnamefont {S.}~\bibnamefont
  {Feng}}, \bibinfo {author} {\bibfnamefont {C.}~\bibnamefont {Kane}}, \bibinfo
  {author} {\bibfnamefont {P.~A.}\ \bibnamefont {Lee}}, \ and\ \bibinfo
  {author} {\bibfnamefont {A.~D.}\ \bibnamefont {Stone}},\ }\href {\doibase
  10.1103/PhysRevLett.61.834} {\bibfield  {journal} {\bibinfo  {journal}
  {Physical Review Letters}\ }\textbf {\bibinfo {volume} {61}},\ \bibinfo
  {pages} {834} (\bibinfo {year} {1988})}\BibitemShut {NoStop}%
\end{thebibliography}%

\end{document}